\documentclass[pra,superscriptaddress,nofootinbib,twocolumn]{revtex4-1}
\usepackage{amsmath,graphicx,epsfig,color,amsfonts,amssymb,bbm}
\usepackage[hypertex]{hyperref}
\usepackage{amsthm}

\newcommand{\ket}[1]{|#1\rangle}
\newcommand{\bra}[1]{\langle#1|}
\newcommand{\braket}[1]{\left\langle #1 \right\rangle}
\newcommand{\one}{\mathbbm{1}}
\newcommand{\zero}{\mathbf{0}}
\renewcommand{\S}{\mathcal{S}}
\newcommand{\Sk}{\S^{(K)}}
\newcommand{\Sq}{\Sk_{\mbox{\tiny qm}}}
\newcommand{\Sqk}[1]{\S^{(#1)}_{\mbox{\tiny qm}}}

\newcommand{\proj}[1]{\left| #1 \right\rangle\!\!\left\langle #1 \right|}

\definecolor{nblue}{rgb}{0.2,0.2,0.7}
\definecolor{ngreen}{rgb}{0.2,0.6,0.2}
\definecolor{nred}{rgb}{0.8,0.2,0.2}
\definecolor{nblack}{rgb}{0,0,0}

\begin{document}

\title{Bell inequalities for three systems and arbitrarily many measurement outcomes}

\author{Basile Grandjean}
\author{Yeong-Cherng~Liang}
\author{Jean-Daniel Bancal}
\affiliation{Group of Applied Physics, University of Geneva, CH-1211 Geneva 4, Switzerland.}
\author{Nicolas Brunner}
\affiliation{H.H. Wills Physics Laboratory, University of
Bristol, Bristol, BS8 1TL, United Kingdom.}
\author{Nicolas Gisin}
\affiliation{Group of Applied Physics, University of Geneva, CH-1211 Geneva 4, Switzerland.}

\date{\today}

\begin{abstract}
We present a family of Bell inequalities for three parties and arbitrarily many outcomes, which can be seen as a natural generalization of the Mermin-Bell inequality. For a small number of outcomes, we verify that our inequalities define facets of the polytope of local correlations. We investigate the quantum violations of these inequalities, in particular with respect to the Hilbert space dimension. We provide strong  evidence that the maximal quantum violation can be reached only using systems with local Hilbert space dimension exceeding the number of measurement outcomes. This suggests that our inequalities can be used as multipartite dimension witnesses.
\end{abstract}

\maketitle

\section{Introduction}

Bell inequalities are constraints on experimental statistics that have to be satisfied by any locally causal theory~\cite{J.S.Bell:Speakable}. In such a theory, the correlations observed between experimental outcomes can be attributed to the common past of the physical systems that give rise to these correlations. Since the pioneering work of Bell~\cite{J.S.Bell:1964}, it has been known that quantum theory predicts correlations that can violate Bell inequalities. Experimental violations of Bell inequalities thus bring out an intriguing aspect of quantum mechanics, challenging us to change our world view at the most fundamental level.

In recent decades, Bell inequalities have also found various applications in quantum information science. Among others, violations of Bell inequalities play a crucial role in the security of quantum cryptographic protocol~\cite{A.K.Ekert:PRL:1991,DIQKD} and  in the generation of trusted random numbers (randomness expansion) \cite{ BIV:Randomness,rand_colbeck}. In addition, they can  be used to witness the dimension of the underlying physical system~\cite{DimWitness,ClDimWitness}. They also find applications in 
self-testing \cite{mayers,tzyh} of quantum devices,
certification of entangled measurements~\cite{Rabello:PRL,TN} and, in some cases, provide nontrivial estimates of the underlying quantum state~\cite{DISE}. More recently, there has been renewed interest in using them to witness (multi-partite) entanglement without resorting to calibrated devices or assumption on the dimension of Hilbert space~\cite{DIEW,K.F.Pal:1102.4320}. For given Hilbert space dimension, there is also the possibility to both lower and upper bound the underlying entanglement based on such quantum violations~\cite{Verstraete:2002,SDIBE}, and to test the structure of multipartite entanglement \cite{sharam}.

It is thus interesting and useful to derive Bell inequalities for a scenario involving an arbitrary number of parties, measurements, and outcomes. In particular it is relevant to look for families of Bell inequalities tailored for systems of arbitrarily high dimensions. While several families have been derived in the bipartite case \cite{CGLMP,BKP}, much less is known in the multipartite case \cite{cavalcanti,salles,SvetCGLMP,chen,GUBI}, for which most studies have focused on the case of binary outcomes (see e.g.~\cite{Mermin,WWZB}).

In this work, we present a family of Bell inequalities for three parties and arbitrarily many measurement outcomes (Sec.~\ref{Sec:BI}). These inequalities are naturally suited for high-dimensional systems, although the nonlocality of such systems  generally also can be tested with Bell inequalities having  fewer outcomes. An appealing feature of our inequalities is that they reduce to well-known and useful Bell inequalities in certain special cases. For the case of binary outcomes, we recover the Mermin-Bell inequality~\cite{Mermin}, while in the bipartite case, we recover the Collins-Gisin-Linden-Massar-Popescu (CGLMP) inequalities~\cite{CGLMP}. Thus our inequalities can be considered as a natural generalization of the Mermin-Bell inequality to a scenario involving an arbitrary number of outcomes. They can also be considered as a tripartite generalization of the CGLMP inequalities. For a small number of outcomes (up to eight) we checked that our inequalities are tight, that is, that they define facets of the polytope of local correlations. 
Quantum violations of our inequalities are discussed in Sec.~\ref{Sec:QuantumViolation}. 
We give strong numerical evidence that the maximal violation of our inequalities requires quantum states of local Hilbert space dimension larger than the number of outcomes, a feature already reported for other Bell inequalities~\cite{DimWitness,TV:2008,TV:2010,Briet:2011}. 
Before concluding in Sec.~\ref{Sec:Conclusion}, we  present other tight tripartite Bell inequalities, in particular giving another possible generalization of the Mermin-Bell inequality to the scenario with three measurement outcomes.

\section{Family of tripartite Bell inequalities}
\label{Sec:BI}

We consider a scenario involving three spatially separated parties (henceforth referred as Alice, Bob and Charlie), and with each of them performing 2 alternative $K$-outcome measurements. We denote by $A_x$, $B_y$, $C_z=0,1,\ldots,K-1$, respectively, the measurement outcome (output) of Alice's $x$-th measurement, Bob's $y$-th measurement and Charlie's $z$-th measurement setting. In these notations, our Bell inequalities read as:
\begin{align}\label{Eq:MerminCGLMP}
	\Sk&=\braket{[A_2-B_1+C_1]_K}+\braket{[A_1+B_2-C_1]_K}\nonumber\\
	&+\braket{[-A_1+B_1+C_2]_K}+\braket{[-A_2-B_2-C_2-1]_K},\nonumber\\
	&\ge K-1
\end{align}
where $[X]_K$ stands for $X$ modulo $K$ and
\begin{gather}
	\braket{[X]_K}\equiv\sum_{j=0}^{K-1}\, j\, P(X=j\,\text{mod}\, K),
\end{gather}
is the bracket notation introduced in Ref.~\cite{Acin06} (see also Ref.~\cite{GUBI}).

Note from Eq.~\eqref{Eq:MerminCGLMP} that $\Sk$ is an expression that is invariant under cyclic permutation of parties, i.e., $A\to B\to C\to A$. To see that inequality~\eqref{Eq:MerminCGLMP} indeed represents a legitimate Bell inequality, i.e., it is a constraint that has to be satisfied by any local correlations, we make use of the following facts:
\begin{enumerate}
	\item It suffices to consider deterministic classical strategies for determining the minimal value of $\Sk$ allowed in a local theory,\footnote{This follows from the fact that the set of local correlations for any finite number of parties, inputs and outputs is a convex polytope of which each extreme point corresponds to a deterministic classical strategy.}
	\item For any integer $X$ and $Y$,  $[X]_K+[Y]_K\ge [X+Y]_K$,
	\item The sum of all the terms in the brackets in Eq.~\eqref{Eq:MerminCGLMP} gives a constant, i.e.,
		\begin{align}
		   [&A_2-B_1+C_1+A_1+B_2-C_1-A_1+B_1+C_2 \nonumber\\
		   -&A_2-B_2-C_2-1]_K=[-1]_K=K-1.
		\end{align}
\end{enumerate}
Moreover, it is also easy to see that the local bound of $K-1$ is always attainable, for example, by setting all the outputs $A_1=A_2=B_1=B_2=C_1=C_2=0$.

For the case of two outputs, i.e., $K=2$, it is straightforward to verify that inequality~\eqref{Eq:MerminCGLMP} is equivalent to the well-known tripartite Mermin-Bell~\cite{Mermin}  inequality
\begin{equation}
	E(2,2,2)-E(2,1,1)-E(1,2,1)-E(1,1,2)\le 2,
\end{equation}
where the correlator is defined as:
\begin{equation}
	E(x,y,z)=\sum_{A_x,B_y,C_z=0,1} (-1)^{A_x+B_y+C_z}P(A_x,B_y,C_z),
\end{equation}
and $P(A_x,B_y,C_z)$ is the joint probability of registering the outcomes $(A_x,B_y,C_z)$ conditioned on Alice, Bob and Charlie, respectively,  performing the $x$-th, $y$-th and $z$-th measurement.

On the other hand, if $C_1=C_2=0$, then applying the relabeling $B_2\to[-B_2]_K$ in inequality \eqref{Eq:MerminCGLMP} gives
\begin{align}\label{Eq:CGLMP}
	&\braket{[A_2-B_1]_K}+\braket{[A_1-B_2]_K}\nonumber\\
	+&\braket{[-A_1+B_1]_K}+\braket{[B_2-A_2-1]_K}\ge K-1,
\end{align}
which is just the CGLMP inequality~\cite{CGLMP} written in the form of brackets~\cite{Acin06}. Hence our inequality reduces to the CGLMP inequality in the bipartite case. Note, however, that inequality~\eqref{Eq:MerminCGLMP} is not a simple lifting~\cite{lifting} of the CGLMP inequality, and hence a genuine tripartite Bell inequality.

Indeed a nice feature of both the Mermin-Bell~\cite{WWZB} and the CGLMP-Bell inequalities~\cite{Masanes:CGLMP} is that they are tight, that is, they represent facets of the local polytope (i.e., a boundary of the set of local correlations with maximal dimension). Here we verified that inequality~\eqref{Eq:MerminCGLMP} for $K\leq8$ represents also a facet of the corresponding local polytope. We conjecture that inequality \eqref{Eq:MerminCGLMP} is indeed facet-defining for all $K\ge 2$.

\section{Quantum violations}
\label{Sec:QuantumViolation}

\subsection{Preliminaries}

Here, we study the quantum violations of inequality \eqref{Eq:MerminCGLMP}. We denote the quantum value of $\Sk$, cf. Eq.~\eqref{Eq:MerminCGLMP}, by $\Sq$ and the $k$-th positive-operator-valued-measure (POVM) element of Alice's $x$-th measurement by $A^k_x$ (likewise $B^k_y$ for Bob, and $C^k_z$ for Charlie). For ease of comparison, we also express the quantum value of $\Sk$ for a specific quantum state $\ket{\psi}\in\mathbb{C}^{d_1}\otimes\mathbb{C}^{d_2}\otimes\mathbb{C}^{d_3}$ and specific measurements --- denoted by $\Sq(\ket{\psi})$ --- in terms of its resistance to white noise. The visibility of $\ket{\psi}$ with respect to a given Bell inequality is then defined as the supremum of $v\in[0,1]$, such that the mixed state
\begin{equation}
	\rho_v=v\proj{\psi}+(1-v)\frac{\one_{D}}{D},
\end{equation}
does not violate the Bell inequality. Here $D=d_1 d_2 d_3$ is the dimension of the Hilbert space of $\ket{\psi}$; $d_j$ is the rank of the reduced density matrix of the $j$-th subsystem, and $\one_D/D$ is the white noise (maximally mixed state) acting on $\mathbb{C}^D$. For inequality~\eqref{Eq:MerminCGLMP}, the visibility of a given state $\ket{\psi}$ is given by
\begin{equation}
	v=\frac{\Sq(\tfrac{\one_D}{D})-(K-1)}{\Sq(\tfrac{\one_D}{D})-\Sq(\ket{\psi})},
\end{equation}
where $\Sq(\tfrac{\one_D}{D})$ is the quantum value for $\one_D/D$ when evaluated using the measurements that give $\Sq(\ket{\psi})$.

\subsection{Binary outputs ($K=2$)}

This case is well studied in the literature: the minimal value of $\Sqk{2}$ is 0 --- which is also the algebraic minimum of the expression $\S^{(K=2)}$ --- and can be achieved using the 3-qubit Greenberger-Horne-Zeilinger (GHZ) state:
\begin{equation}\label{Eq:GHZ2}
	\ket{\text{GHZ}_2}= \frac{1}{\sqrt{2}} \left( \ket{000} + \ket{111}  \right),
\end{equation}
in conjunction with rank-1 projective measurements. This quantum strategy also corresponds to the so-called GHZ paradox~\cite{GHZParadox}, and gives a visibility of $v=1/2$. 

\subsection{Ternary outputs ($K=3$)}

\subsubsection{Bounded Hilbert space dimension}

By optimizing numerically over the most general POVMs\footnote{Here, we employ the iterative algorithm of Ref.~\cite{Liang:PRA:2009} (see also Refs.~\cite{K.F.Pal:2009,Y.C.Liang:PRA:042103}) via YALMIP~\cite{YALMIP} to perform the numerical optimization.} up to local Hilbert space dimension of $d_1=d_2=d_3=3$,  the best quantum violation we have found is $\Sqk{3}(\ket{\psi_3})=\frac{7-3 \sqrt{3}}{2} \approx 0.9019$. This is achieved using the 3-qubit state

\begin{equation}\label{Eq:OptState:3output}
\ket{\psi_3}= \frac{(3-2\sqrt{3})\ket{000} + \ket{011} + \ket{101} + \ket{110} }{2 \sqrt{6-3 \sqrt{3}}}
\end{equation}
and the following measurements:
\begin{gather}
	A_{1}^k=
	\left\{ \frac{\one_2+\sigma_{z}  }{2}, \frac{\one_2-\sigma_{z}  }{2}, \zero \right\},\,
	A_{2}^k=
	\left\{ \frac{\one_2-\sigma_{x}  }{2}, \zero, \frac{\one_2+\sigma_{x}  }{2} \right\}, \nonumber\\
	B_{1}^k=
	\left\{ \frac{\one_2 - \sigma_{z}  }{2}, \frac{\one_2 + \sigma_{z}  }{2}, \zero \right\},\,
	B_2^k=
	\left\{  \frac{\one_2 + \sigma_x  }{2},  \zero, \frac{\one_2 - \sigma_x  }{2}\right\},\nonumber\\
	C_{1,2}^k=
	\left\{ \frac{\one_2 - \sigma_{x,z}  }{2},  \frac{\one_2 + \sigma_{x,z}  }{2}, \zero \right\},
\end{gather}
where $\sigma_{x}=\ket{0}\bra{1}+\ket{1}\bra{0}$, $\sigma_{z}=\proj{0}-\proj{1}$ are the Pauli $x$ and $z$ matrices, $\zero$ is the $2\times2$ zero matrix, and the POVM elements are ordered in the bracket with increasing value of $k$. The above quantum violation translates into a visibility of $v=65.26\%$.

Interestingly, by increasing the local Hilbert space dimension $d_j$, we find a gradually stronger violation [i.e., a smaller value of $\Sqk{3}$] that reaches the minimal value of  $\Sqk{3}\approx 0.9005$, corresponding to a visibility of $v=64.04\%$. Considering states of local dimension up to 10, the state $\ket{\psi_3^*}$ giving the best violation has local dimensions $d_1=d_2=6, d_3=2$. The measurements are rank 1 projectors. Table~\ref{Tbl:3Output-general} summarizes the results for states of bounded local Hilbert dimension.

It is natural to ask whether this violation represents the optimal quantum value. Using a converging hierarchy of semidefinite programs~\cite{QMP.Hierarchy1,QMP.Hierarchy2,QMP.Hierarchy3} (up to a partial list of 5th order operators), we found the lower bound $\Sqk{3}\geq0.8507$, which leaves a small gap. Hence it could be the case that stronger quantum violations are possible. Also, using a more refined implementation of these semidefinite programs, it can also be shown\footnote{Within a numerical precision of $10^{-7}$.} that $\Sqk{3}(\ket{\psi_3})=\frac{7-3 \sqrt{3}}{2} \approx 0.9019$ is indeed the strongest possible quantum violation if both the state and the POVMs are invariant under an arbitrary permutation of parties.

\begin{table}[t!]
	\begin{ruledtabular}
        \begin{tabular}{|c||c|c|c|c|c|}\hline
        $d $ & 2& 3 & 4 & 5 & 6 \\ \hline
        $\Sqk{3}$ & 0.9019 & 0.9019 & 0.9019\footnote{The violation in this case is actually stronger than the one with $d\le3$, but the difference is less than $10^{-4}$.} 
        & 0.9015  & 0.9005 \\                  \hline	
        $[d_1;d_2;d_3]$ & [2;2;2] & [2;2;2] & [4;4;2] & [5;5;2]  & [6;6;2] \\ \hline
	\end{tabular}
        \end{ruledtabular}
	\caption{\label{Tbl:3Output-general} Best quantum violations $\Sqk{3}$ for bounded local Hilbert space dimension. From top to bottom, we have, respectively, the bound on the local Hilbert space dimension (i.e., $d_j\le d$), the best quantum violation found and the ranks of the reduced density matrices of the corresponding optimal quantum state. Note that the best violation $\Sqk{3}$ is found for $d=6$. For even higher Hilbert space dimensions (up to $d= 10$), no improvement was found.}
\end{table}

\subsubsection{Specific states and measurements}

We first investigate the violation of our inequality for well-known states, such as $\ket{\text{GHZ}_2}$, the 3-qutrit GHZ state,
\begin{equation}
\ket{\text{GHZ}_3}= \frac{1}{\sqrt{3}} \left( \ket{000} +\ket{111} + \ket{222} \right),
\end{equation}
the $W$ state $\ket{W}$,
 \begin{equation}
\ket{W}= \frac{1}{\sqrt{3}} \left( \ket{001} +\ket{010} + \ket{100} \right)
\end{equation}
and the fully antisymmetric 3-qutrit state (also known as the Aharonov state)
\begin{equation}
\ket{\mathcal{A}}=\frac{1}{\sqrt{6}} \left( \ket{012} + \ket{120} + \ket{201} -\ket{021} -\ket{102} - \ket{210} \right).
\end{equation}
The numerical optimization results for these states are summarized in Table~\ref{Tbl:3Output}. Again, we see that qubit states seem to achieve stronger violations compared to genuine qutrit states, even though the inequality involves three measurement outcomes.

Second, we investigate the violation of our inequality under specific measurements, In particular, we consider Fourier-transformed measurements, known to be optimal in the case of CGLMP inequalities~\cite{acin02, Kaslikowski:PRL:00} and some of its generalizations~\cite{SvetCGLMP}. Here, following the method of Ref.~\cite{S.L.Braunstein:PRL:1992} based on the Bell operator, we find the largest violation to be $\sim 1.206$ for a 3-qutrit state.

\begin{table}[t!]
\begin{ruledtabular}
\begin{tabular}{|c|cc|}
Quantum state $\ket{\psi}$                & $\Sqk{3}(\ket{\psi}) $ & $v_{\ket{\psi}}$  \\
\hline
$\ket{\psi_3^*}$     & 0.9005                    & 0.6404 \\
$\ket{\psi_3}$     & 0.9019                    & 0.6456 \\
$\ket{W}$     & 1.2249                    & 0.7075  \\
$\ket{\text{GHZ}_2}$ & 1.2500                    & 0.7143 \\
$\ket{\text{GHZ}_3}$ & 1.3333                    &  0.7500 \\
$\ket{\mathcal{A}}$     &  1.4508                    & 0.7846 \\
\end{tabular}
\caption{\label{Tbl:3Output} Summary of the quantum violation of inequality~\eqref{Eq:MerminCGLMP} with $K=3$ for various states. The second column gives the best (i.e., lowest) quantum value found for the respective state and the third column gives the corresponding visibility.}
\end{ruledtabular}
\end{table}

\subsection{More than three outputs ($K>3$)}

Since the Mermin-Bell inequality is tailored for the GHZ state, which reaches the algebraic bound of the inequality, it is natural to investigate the violation of our inequality 
for $K>3$ by the generalized GHZ state:
\begin{equation}
	\ket{\text{GHZ}_K}= \frac{1}{\sqrt{K}} \sum_{j=0}^{K-1} \ket{j\,j\,j}.
\end{equation}
For $K\leq 10$, the best violation (except for $K=3,7$, see below) is simply given by an integer value:
\begin{equation}
	\Sq(\ket{\text{GHZ}_K})=\left\lfloor\frac{K-1}{2} \right\rfloor,
\end{equation}
which corresponds to the following visibility:
\begin{equation}\label{Eq:v-GHZ}
	v_{\ket{\text{GHZ}_K}}=\frac{K-1}{2(K-1)-\left\lfloor\frac{K-1}{2} \right\rfloor}.
\end{equation}
Note that this threshold visibility converges to $\tfrac{2}{3}$ for asymptotically large $K$. 

It turns out that in two case, $K=3$ and $K=7$, larger violations could be found. Interestingly, for the case $K=7$, a small improvement (2.9907 compared to 3) can be obtained using genuine POVMs (i.e. non-projective measurements). This stronger violation, however, leads to a weaker resistance to white noise [0.6708 compared to $v_{\ket{\text{GHZ}_K}}=\tfrac{2}{3}$ from Eq.~\eqref{Eq:v-GHZ}]. This is astonishing in the sense that the strength of violation and the robustness to white noise are generally considered to relate monotonously to each other. But as we see here, if genuine POVMs (or higher-rank projectors) are used, their effect on white noise is no longer trivial and hence the monotonicity is not necessarily preserved. It would be interesting to see if this non-monotonicity could have practical implication in the choice of measurements for tasks based on quantum nonlocality.

Next we investigated the violation of inequality~\eqref{Eq:MerminCGLMP} achievable with local Hilbert space dimension $d_j\le K$.
Similarly to the case $K=3$, we employed the iterative algorithm of Refs.~\cite{Liang:PRA:2009,K.F.Pal:2009} and the technique of the Bell operator~\cite{S.L.Braunstein:PRL:1992}. The results, summarized in Table~\ref{Tbl:Quantum}, indicate that the generalized GHZ state is not optimal for $K>2$. Also for increasing values of $K$, it seems that the best threshold visibility found here does not follow any regular pattern, in contrast with the CGLMP inequalities.

Finally it is interesting to note the structure of the optimal state (for $d_j\le K$), which seems to always involve a qubit for one the parties.

\begin{table}[t!]
\begin{ruledtabular}
\begin{tabular}{r|cc|cc|cc}
$K$ &  $\Sq(\ket{\text{GHZ$_K$}})$ & $v_{\ket{\text{GHZ}_K}}$
& $\Sq(\ket{\psi^*_K})$ & $v_{\mbox{\tiny{qm}}}$ & $[d_1;d_2;d_3]$
\\ \hline
  2 &   0  & 0.5 & 0 & 0.5  & $[2;2;2]$\\
  3 &   1.3333  & 0.7500 & 0.9005 & 0.6404 & $[6;6;2]$\\
  4 &   1.0000  & 0.6000 & 0.7657  &  0.5731 & $[4;4;2]$ \\
  5 &   2.0000  & 0.6667 & 1.4763 &  0.5691 & $[3;3;2]$  \\
  6 &   2.0000  & 0.6250 & 1.4652 &  0.5858 & $[6;6;2]$ \\
  7 &   2.9907  & 0.6708  & 2.2924 & 0.6095 &  $[7;7;2]$ \\
\end{tabular}
\caption{\label{Tbl:Quantum} Summary of the quantum violations for inequality~\eqref{Eq:MerminCGLMP}. The second and the third column gives the violation for $\ket{\text{GHZ}_K}$ and the corresponding visibility. The next two columns give the largest quantum violation for states such that $d_j\le K$; the optimal state is denoted $\ket{\psi^*_K}$. The ranks of the reduced density matrices of $\ket{\psi^*_K}$ are given in the last column.}
\end{ruledtabular}
\end{table}

\section{Other Bell inequalities}
\label{Sec:OtherBI}

Given that inequality~\eqref{Eq:MerminCGLMP} is only symmetrical with respect to cyclic permutations of parties, one may wonder whether there exists a more symmetrical generalization of the tripartite Mermin-Bell inequality to more outputs that still preserves the facet-defining property. To answer this question, we made use of the techniques discussed in Ref.~\cite{JD:SymmIneq} and computed all tripartite, two-input, three-output facet-defining Bell inequalities that are symmetrical with respect to arbitrary permutations of parties and which depend only on the sum of the outputs (see Appendix~\ref{App:Symm323} for details). Among these inequalities, we found a candidate that also reduces to the tripartite Mermin-Bell inequality for $K=2$ outputs:
\begin{equation}\label{Eq:Mermin2}
\begin{split}
&\braket{[A_2+B_2+C_2]_K}+\braket{[A_2+B_2+C_2+1]_K}\\
+&\braket{[A_2+B_1+C_1]_K}+\braket{[A_2+B_1+C_1+1]_K}\\
-&3\braket{[A_1+B_1+C_1]_K}-2\braket{[A_1+B_1+C_1+1]_K}\\
+&\braket{[A_2+B_2+C_1]_K}+\circlearrowright \,\ge 2,\qquad K=2,3
\end{split}
\end{equation}
where $\circlearrowright$ means terms (brackets) that must be added to ensure that the expression is symmetrical with respect to arbitrary permutations of parties. Unfortunately, the facet-defining property of the inequality is not preserved for $K=4$ and 5.

The best quantum violation that we have found for inequality~\eqref{Eq:Mermin2} with $K=3$ is -0.1920, corresponding to a visibility of $64.60\%$, using a 3-qutrit state.

On a separate note, while searching for another possible generalization of the Mermin-Bell inequality to more outputs, we found a possible generalization of Sliwa's 7th inequality \cite{Sliwa} to an arbitrary number of outputs (see Appendix~\ref{App:Sliwa} for details).

\section{Concluding Remarks}
\label{Sec:Conclusion}

We have presented a simple family of Bell inequalities for tripartite systems and  arbitrarily many measurement outcomes, shown them to be tight for a small number of outputs, and investigated their violations in quantum mechanics. 

These inequalities serve as a natural generalization of both the Mermin-Bell inequality and the CGLMP inequalities. 
Apparently, however, these inequalities do not preserve some of the nice features of the latter inequalities. In particular, they do not seem to become more resistant to white noise (characterized in terms of visibility) as the number of measurement outcomes increases, contrary to the CGLMP inequalities. Also, it seems that our inequalities cannot be readily generalized to the case of four parties and more. In fact, we performed an exhaustive search for similar 4-partite (facet-defining) inequalities that are invariant with respect to cyclic permutations of parties, but to no avail.

Nonetheless, there are also features of the present inequalities that deserve further investigation. Firstly, all quantum states that violate the $K$-outcome inequality maximally do not  seem to require a tripartite quantum state that has local Hilbert space dimensions $d_1=d_2=d_3=K$. In particular, one of them is always a qubit while the other two systems have local dimension equal or greater than the number of outcomes. 
In accordance with recent work~\cite{TV:2008,DimWitness,TV:2010,Briet:2011} this shows that quantum systems of local Hilbert space dimension exceeding the number of measurement outcomes can be required in order to reach the maximal violation of a Bell inequality.
This suggest that these inequalities could be used as tripartite dimension witnesses~\cite{DimWitness}.

\emph{Note added.} Recently, we became aware of a related work by Arnault~\cite{arnault}. We note, however, that our inequalities do not follow directly from the inequalities presented in Ref.~\cite{arnault} since the latter involve expectation values of the product of observables from the same party (and are thus better seen as examples of noncontextual inequalities~\cite{Liang:PR:2011}).

\begin{acknowledgments}
YCL acknowledges useful discussion with Tam\'as V\'ertesi. This work is supported by the UK EPSRC, the Swiss NCCR ``Quantum Science and Technology",  the CHIST-ERA DIQIP, and the European ERC-AG QORE.
\end{acknowledgments}

\appendix

\section{Complete list of symmetric, tripartite, two-input, three-output Bell inequalities involving only the sum of outputs }
\label{App:Symm323}

There are 9 inequivalent classes of facet-defining tripartite, two-input, three-output Bell inequalities which are symmetrical with respect to arbitrary permutations of parties and that involve only the sum of the outputs. For convenience, let us define
\begin{equation*}
E(j|xyz)=P(A_x+B_y+C_z=j\,\text{mod}\, 3).
\end{equation*}
Note that $[A_x+B_y+C_z]_3=\sum_{j=0}^{2} j E(j|xyz)$. With these notations, the 9 inequalities read as:
\begin{widetext}
\begin{align}
\label{Eq:Symm1}
E(2|000) - E(1|001) - E(1|011) - E(2|011) + 2E(2|111) + \circlearrowright \le  0\\
- 2E(1|000) - E(1|001) - 2E(1|011) + 2E(1|111) + \circlearrowright \le  0\\
- 8E(1|000) - 2E(2|000) - E(1|001) + 2E(2|001) - 2E(1|011) - 2E(2|011) + 2E(1|111) - E(2|111) + \circlearrowright \le  0\\
- 2E(1|000) - E(2|000) - E(1|001) - 2E(2|001) - 2E(1|011) - E(2|011) + 5E(1|111) + 4E(2|111) + \circlearrowright \le  0\\ \label{Eq:Symm4}
- 2E(1|000) - 2E(2|000) - E(1|001) - E(2|001) - 2E(1|011) - 2E(2|011) + 5E(1|111) + 5E(2|111) + \circlearrowright \le  0\\
- E(1|000) - E(1|001) - E(2|001) - 3E(1|011) - E(2|011) + 4E(1|111) + 3E(2|111) +  \circlearrowright \le  0\\
- 3E(1|000) - E(2|000) - E(1|001) - E(2|001) - E(1|011) + 3E(1|111) + E(2|111) + \circlearrowright \le  0\\
- 6E(1|000) - 3E(2|000) - E(1|001) - 2E(2|001) - E(1|011) + E(2|011) + 3E(1|111) + \circlearrowright \le  0\\
\label{Eq:Symm9}
- 3E(1|000) + E(2|000) - E(1|001) - 4E(1|011) - E(2|011) + 3E(1|111) + 2E(2|111) + \circlearrowright  \le  0
\end{align}
\end{widetext}
where we have used $\circlearrowright$ to denote missing terms which must be added to ensure that the inequality is symmetrical with respect to arbitrary permutation of parties.

The symmetrical generalization of the Mermin-Bell inequality given in Eq.~\eqref{Eq:Mermin2} was obtained from inequality~\eqref{Eq:Symm4} above by expressing $E(j|x,y,z)$ in terms of brackets $\braket{[A_x+B_y+C_z+l]_K}$ where $l=0,1$.

It is worth noting that, numerically, we have found inequality~\eqref{Eq:Symm1} and \eqref{Eq:Symm9} to give better tolerance to white noise than inequality~\eqref{Eq:MerminCGLMP}.

\section{A generalization of Sliwa's 7th inequality to an arbitrary number of outputs}
\label{App:Sliwa}

A possible symmetric generalization of Sliwa's 7th inequality \cite{Sliwa} to an arbitrary number of outputs reads as:
\begin{equation}\label{Eq:symmineq}
\begin{split}
&2 \braket{[A_1+B_1+C_1]_K}    +2\braket{[-A_1-B_1-C_1-1]_K} \\ 
+&\braket{[-A_1-B_1-C_1]_K}+3\braket{[-A_2-B_2-C_2-1]_K}\\ 
+&\braket{[A_2+B_2+C_2-1]_K} +\braket{[A_2+B_2+C_2]_K} \\ 
+&\braket{[-A_2+B_1+C_1]_K }  + \braket{[-A_1+B_2+C_2]_K }+\circlearrowright\\
  \ge &\quad 6(K-1),
\end{split}
\end{equation}
where the local bound of $6(K-1)$  deduced numerically has been verified for $K\leq 8$; it is clear that, by setting $A_x=B_y=C_z=0$ for $x,y,z=1,2$, the bound $6(K-1)$ can always be attained. For the same range of $K$, we have also verified that inequality~\eqref{Eq:symmineq} represents a  a facet of the corresponding local polytope. We conjecture that both the local bound and the facet-defining property of inequality~\eqref{Eq:symmineq} hold for general $K$.

We have also investigated the quantum violations of inequality~\eqref{Eq:symmineq} (see Table \ref{Tbl:SLIWA}).

\begin{table}[t!]
\begin{ruledtabular}
\begin{tabular}{l|c|cc|c}
$d=K$ & Local Bound & $\S_{\rm qm}$  & $v_{qm}$  & $[d_1;d_2; d_3]$\\
\hline
2 & 6  &  4.6667 & 0.6000     & [2;2;2]\\
3 & 12 & 9.8079  & 0.6460    & [3;3;3]\\
4 & 18 & 14.7913 & 0.6516   & [4;4;4]\\
5 & 24 & 19.6829 & 0.6495   & [5;5;5]\\
6 & 30 & 24.5107 & 0.6456   & [6;6;6]
\end{tabular}
\caption{\label{Tbl:SLIWA} Best quantum violation and corresponding visibility found for inequality~\eqref{Eq:symmineq}. The rank of the reduced density matrices of the optimal state found is given in the last column.}
\end{ruledtabular}
\end{table}


\begin{thebibliography}{99}

\bibitem{J.S.Bell:Speakable} J.~S.~Bell, {\em Speakable and
    Unspeakable in Quantum Mechanics} (Cambridge University
    Press, Cambridge, 2004).

\bibitem{J.S.Bell:1964} J.~S. Bell, Physics (Long Island City,
    N.Y.)  {\bf 1}, 195 (1964).

\bibitem{A.K.Ekert:PRL:1991} A.~K.~Ekert, Phys. Rev. Lett. {\bf 67}, 661
    (1991).

\bibitem{DIQKD} J. Barrett, L. Hardy, A. Kent, Phys. Rev. Lett. {\bf 95}, 010503 (2005); A.~Ac\'in, N.~Brunner, N.~Gisin, S.~Massar, S.~Pironio, and V.~Scarani, \prl { \bf 98}, 230501 (2007).

\bibitem{BIV:Randomness} S.~Pironio, A.~Ac\'in, S.~Massar, A. Boyer~de~la~Giroday, D.~N.~Matsukevich, P.~Maunz, S.~Olmschenk, D.~Hayes, L.~Luo, T.~A.~Manning and C. Monroe, Nature (London) {\bf 464}, 1021 (2010).

\bibitem{rand_colbeck} R. Colbeck and A. Kent, Journal of Physics A: Mathematical and Theoretical {\bf 44}, 095305 (2011).

\bibitem{DimWitness} N.~Brunner, S.Pironio, A.~Ac\'in, N.~Gisin, A.~A.~M\'ethot, and V.~Scarani, Phys. Rev. Lett. {\bf 100}, 210503 (2008).

\bibitem{ClDimWitness} R.~Gallego, N.~Brunner, C.~Hadley, and A.~Ac\'in, Phys. Rev. Lett. {\bf 105}, 230501 (2010).

\bibitem{mayers} D. Mayers and A. Yao. Quantum Inform. Comput. {\bf 4}, 273 (2004).

\bibitem{tzyh} M. McKague, T.H. Yang, and V. Scarani,  arXiv:1203.2976.

\bibitem{Rabello:PRL} R. Rabelo,  M. Ho, D. Cavalcanti, N. Brunner and V. Scarani, Phys. Rev. Lett. {\bf 107}, 050502 (2011).

\bibitem{TN} T.~Vertesi and M.~Navascu\'es, Phys. Rev. A {\bf 83}, 062112 (2011).

\bibitem{DISE} C.-E.~Bardyn, T.~C.~H.~Liew, S.~Massar, M.~McKague, and V.~Scarani, \pra { \bf 80}, 062327 (2009).

\bibitem{DIEW} J.-D.~Bancal, N.~Gisin, Y.-C.~Liang and S.~Pironio, Phys. Rev. Lett. {\bf 106}, 250404 (2011).

\bibitem{K.F.Pal:1102.4320} K.~F.~P\'al and T.~V\'ertesi, Phys. Rev. A {\bf 83}, 062123 (2011).

\bibitem{Verstraete:2002} F.~Verstraete and M. M.~Wolf, Phys.Rev. Lett. {\bf 89}, 170401 (2002).

\bibitem{SDIBE} Y-C.~Liang, T.~V\'ertesi, and N.~Brunner, Phys. Rev. A {\bf 83}, 022108 (2011).

\bibitem{sharam} N. Brunner, J. Sharam, and T.~V\'ertesi, Phys. Rev. Lett. {\bf 108}, 110501 (2012).


\bibitem{CGLMP} D. Collins, N. Gisin, N. Linden, S. Massar, and S. Popescu, Phys. Rev. Lett. {\bf 88}, 040404 (2002).

\bibitem{BKP} J.~Barrett, A.~Kent, and S.~Pironio, Phys. Rev. Lett. {\bf 97}, 170409 (2006).

\bibitem{cavalcanti} E.G. Cavalcanti, C.J. Foster, M.D. Reid, P.D. Drummond, Phys. Rev. Lett. {\bf 99}, 210405 (2007).

\bibitem{salles} A. Salles, D. Cavalcanti, A. Acin, D. P\'erez-Garcia, M.M. Wolf, Quant. Inf. Comp. {\bf 10}, 0703-0719 (2010).

\bibitem{SvetCGLMP} J.-D. Bancal, N.~Brunner, N.~Gisin, and Y.-C. Liang, Phys. Rev. Lett. {\bf 106}, 020405 (2011).

\bibitem{chen} J.-L. Chen, D.-L. Deng, H.-Y. Su, C. Wu, C.H. Oh, Phys. Rev. A {\bf 83}, 022316 (2011).

\bibitem{GUBI} J.-D. Bancal, C. Branciard, N. Brunner, N. Gisin, Y.-C. Liang, J. Phys. A: Math. Theor. {\bf 45}, 125301 (2012).


\bibitem{Mermin} N.~D.~Mermin, Phys. Rev. Lett. {\bf 65}, 1838 (1990).

\bibitem{WWZB} R. F. Werner and M. M. Wolf, Phys. Rev. A {\bf 64}, 032112 (2001);  M. \.{Z}ukowski and \v{C}. Brukner, Phys. Rev. Lett. {\bf 88}, 210401 (2002).

\bibitem{TV:2008} T. V\'ertesi and K.~F.~P\'al, Phys. Rev. A {\bf 77}, 042106 (2008);
K.~F.~P\'al and T. V\'ertesi, Phys. Rev. A {\bf 77}, 042105 (2008).

\bibitem{TV:2010} K.~F.~P\'al and T.~V\'ertesi, Phys. Rev. A {\bf 82}, 022116 (2010).

\bibitem{Briet:2011} J. Briet, H. Buhrman, and B. Toner, Comm. Math. Phys. {\bf 305}, 827 (2011).


\bibitem{Acin06} A. Ac\'in, R. Gill, and N. Gisin, Phys. Rev. Lett.  {\bf 95}, 210402 (2005).


\bibitem{lifting} S.~Pironio, J. Math. Phys. {\bf 46}, 062112 (2005).


\bibitem{Masanes:CGLMP} Ll. Masanes, Quantum Inf. Comput. {\bf 3}, 345 (2003).

\bibitem{GHZParadox} D.~M.~Greenberger, M.~A.~Horne, A.~Zeilinger, in ``Bell's Theorem, Quantum Theory, and Conceptions of the Universe", edited by M. Kafatos (Kluwer, Dordrecht, 1989), 69-72; N.~D.~Mermin, \prl { \bf 65}, 3373 (1990).

\bibitem{Liang:PRA:2009} Y.-C.~Liang, C.~W.~Lim, and D.~L.~Deng, Phys. Rev. A {\bf 80}, 052116 (2009).

\bibitem{K.F.Pal:2009} K. F. P\'al and T. V\'ertesi, Phys. Rev. A {\bf 79}, 022120 (2009).

\bibitem{Y.C.Liang:PRA:042103} Y.-C.~Liang and A.~C.~Doherty,
    Phys. Rev. A {\bf 75}, 042103 (2007).

\bibitem{YALMIP} YALMIP : A Toolbox for Modeling and Optimization in MATLAB. J. L\"ofberg in {\em Proceedings of the CACSD Conference}, (IEEE, Taipei, Taiwan, 2004).

\bibitem{QMP.Hierarchy1} M. Navascu\'es, S. Pironio, and    A.~Ac\'in, \prl { \bf
    98}, 010401 (2007).

\bibitem{QMP.Hierarchy2}  M. Navascu\'es, S. Pironio, and   A.~Ac\'in, New J.~Phys., {\bf 10}, 073013 (2008);
    A.~C.~Doherty, Y.-C.~Liang, B.~Toner, and S.~Wehner in  {\em Proceedings of
    the 23rd IEEE Conference on Computational Complexity} (IEEE Computer Society, College Park, MD, 2008), pp. 199-210.

\bibitem{QMP.Hierarchy3} S. Pironio, M. Navascu\'es, A. Ac\'in, SIAM J. Optim. vol \textbf{20}, issue 5, 2157 (2010).

\bibitem{acin02}
A. Ac\'in, T. Durt, N. Gisin, and J.~I.~Latorre, Phys. Rev. A {\bf 65} 052325 (2002).

\bibitem{Kaslikowski:PRL:00}  D. Kaszlikowski, P.~Gnacinski, M. \.{Z}ukowski, W.~Miklaszewski, and A.~Zeilinger, Phys.~Rev.~Lett.  {\bf 85}, 4418   (2000).

\bibitem{S.L.Braunstein:PRL:1992} S.~L.~Braunstein, A.~Mann, and
  M.~Revzen, \prl { \bf 68}, 3259 (1992).

\bibitem{JD:SymmIneq} J.-D.~Bancal, N.~Gisin, and S.~Pironio, J. Phys. A: Math. Theor. {\bf 43} 385303 (2010).

\bibitem{Sliwa} C. Sliwa, Phys. Lett. A {\bf 317}, 165 (2003).

\bibitem{arnault} F. Arnault, arXiv:1107.2255.

\bibitem{Liang:PR:2011} Y.-C. Liang, R. W. Spekkens, and H. M. Wiseman, Phys. Rep. {\bf 506}, 1 (2011).

\end{thebibliography}
\end{document}